\documentclass[12pt]{article}
\usepackage{diagbox}
\usepackage{cite}
\usepackage{amsmath,amssymb,amsfonts}
\usepackage{algorithmic}
\usepackage{caption}
\usepackage{subcaption}
\usepackage{graphicx}
\usepackage{textcomp}
\usepackage{inputenc}
\usepackage[table]{xcolor} 
\usepackage[symbol]{footmisc}

\usepackage[section]{placeins}
\usepackage{hyperref}
\newcolumntype{P}[1]{>{\centering\arraybackslash}p{#1}}
\newcolumntype{M}[1]{>{\centering\arraybackslash}m{#1}}
\def\BibTeX{{\rm B\kern-.05em{\sc i\kern-.025em b}\kern-.08em
    T\kern-.1667em\lower.7ex\hbox{E}\kern-.125emX}}

\begin{document}

\def\spacingset#1{\renewcommand{\baselinestretch}%
{#1}\small\normalsize} \spacingset{1}

  \begin{center}
    {\Large\bf Post-Covid learning assessment of school children:  \\ A Project by CRY \& RILM across four states}\\

   {\large Anushka De\footnote[1]{I thank \href{https://www.cry.org/}{CRY-Child Rights and You} for the data used in this paper.} \\ Indian Statistical Institute, Kolkata\\anushka.isical@gmail.com}
\end{center}

\bigskip
\begin{abstract}
The COVID-19 pandemic has struck education system around the globe. The pandemic initiated an immediate and complete lockdown of all the educational institutions, to maintain social distancing. CRY (Child Rights and You) in collaboration with RILM (Rotary India Literacy Mission) initiated a project to assess the learning abilities of 4000 children across four states: Jammu \& Kashmir, Jharkhand, Manipur and West Bengal. Every child was provided with the books of appropriate
class according to their age in order to test their competency in reading and basic calculation
as well and thereafter the compatible class was determined. The assessments were carried  over 3 quarters in 4 subjects: oral assessments in $1^{st}$ Language, $2^{nd}$
Language and  Mathematics and a writing assessment and a binary variable for improvement/no improvement (1/0) was provided. This paper suggests a measure which gives a unique score for improvement level of students with varied class lag since it will not be a desirable idea to grade the students with varied class lags on the same basis. This paper also investigates and evaluates the progression of student performance over the 3 quarters, suggests the use of a comprehensive score measure for summarising the inter-quarter performance. The analysis of progression has been carried out by gender and state level for male-female and inter state comparison respectively.  

\end{abstract}

%\maketitle
\section{Introduction}
\label{sec:introduction}
The global disruption to education caused by the COVID-19 pandemic is without parallel, and its effects on learning have been severe. See \cite{Donnelly} and \cite{Engzell} for a few early empirical stuides. The crisis brought education systems across the world to a halt, with school closures affecting more than 1.6 billion learners.
School closure due to the pandemic has led to a complete disconnect from education for the vast majority of children or inadequate alternatives like a community-based classes or poor alternatives in the form of online education, including mobile phone-based learning. The risk of dropout among students increased
more than threefold \cite{UNESCO_report}. 

\textbf{CRY (Child Rights and You)} in collaboration with \textbf{RILM (Rotary India Literacy Mission)} is implementing a
project for child development in selected CRY implementation operational areas and the findings reported in this paper was used as the baseline analysis for the same. 
The study assesses the learning level among children in the post covid situation in the states covered
4000 children across 4 states. It focused on the assessment of learning level, compatibility
level, age appropriation class, and dropout status of the students. The study involved students from ages
7-14 years. These issues for each grade were chosen because these play a major role across all subsequent
learning – across subjects – and so the loss of any one of these would have very serious consequences on
all further learning.
This paper reports the analysis of the post pandemic progressive performance of the students.

\section{Methodology and Data}

The study was conducted with 4000 children covered 4 districts across 4 states – Jammu \& Kashmir, West Bengal, Jharkhand, and Manipur.
The steps involved in the assessment are: 
\begin{itemize}
    \item The assessment of the children was done on the basis of three indicators: the status of children- dropouts ( a drop-out is a person who leaves school, university,
etc. before finishing his/ her studies), never enrolled, and laggards, learning as per the age-appropriate class and reading and basic calculation competency as per the age appropriate class
\item To determine the status and to assess the learning, every child was provided with the books of appropriate
class according to their age in order to test their competency in reading and basic calculation
as well.
\item As per the ability of each child to read and calculate, they were scored by the teachers determining
their learning level.
\item There were 50 teachers/assessors working on the field, as part of the assessment. All of them received
training from the RILM (Rotary India Literacy Mission).
\item Finally, the data for all 4000 children were compiled according to their class, determined by their age
and level of learning.
\end{itemize}

\subsection{Demography}

The total number of children screened was 4000, and the age group covered was 7-14 years. Of the 4000 students of which 1734 (43.35\%) are in the centres of West Bengal, 1001 (25\%) in Manipur, 700 (17.5\%) in Jharkhand and 565 (14.12\%) in Jammu \& Kashmir. 
The total male sample accounted for 47.1\% of the total sample whereas the female accounted for 52.9\%.
The children came from 40 different villages across the four states. The total number of Panchayats covered
was 23 in number, Blocks covered were 4, and districts were 4 as well, precisely one from each state.
\begin{table}[h]
    \centering
     \rowcolors{2}{blue!12}{white}
    \begin{tabular}{|M{2cm}|M{2cm}|M{2.5cm}|M{2.2cm}|M{2cm}|M{2cm}|}
 \hline
 \rowcolor{blue!23}
 States & Districts & Blocks & Number of Panchayats & Number of centres & Number of children \\ [2em]
 \hline
 
Jammu \& Kashmir & Bandipore & Sambal, Bandipora,
Hajin & 9 & 9 & 565 \\
 \hline
West Bengal & Jalpaiguri & Banarhat & 4& 15 &1735\\
\hline
Jharkhand & Koderma& Satgawa& 2& 6 &700\\
\hline 
Manipur & Imphal west& Haorang Sabal,Patsoi &
8 &9 &1000\\
 \hline
    \end{tabular}
    \caption{Table showing geographical coverage of the study}     
    \label{geographical_coverage}
\end{table}

\subsection{Data Description}

CRY provided us with the assessments records of the 4000 children for 3 quarters. For each quarter the dataset included Child-ID, CRY Center, State, Sex, Age	Appropriate Class, Compatible Class, Attendance and a binary variable for improvement/no improvement (1/0) for assessments in each of the 4 subjects:oral assessments in $1^{st}$ Language, $2^{nd}$, 
Language and  Mathematics and a writing assessment.

\section{Analysis of Performance}

In this section we provide description of the methods used by us for analysing the performance (in terms of improvement) over the quarters. So an improvement level 4 means that the student showed improvement in all 4 subjects. 

\subsection{Comparison of Performance with level of Class Lag}

The difference between the age-appropriate class and compatible class (class lag) for every student and the improvement level in total number of subjects, is used for cross tabulation.  The number of students lying in every class lag-improvement category is computed in order to compare the level of improvement and degree of class lag.  
The students for whom compatible class is higher than age appropriate class have been excluded.
\begin{table}[h]
    \centering
     \rowcolors{2}{blue!12}{white}
\begin{tabular}{|c|c|c|c|c|c|}\hline
\backslashbox{Class lag}{Improvement}
&\makebox[1.5em]{0}&\makebox[1.5em]{1}&\makebox[1.5em]{2}
&\makebox[1.5em]{3}&\makebox[1.5em]{4}\\\hline\hline
-7&	2&	2	&21&	36&	3\\\hline
-6	&5&	2	&49&	129&	26\\\hline
-5	&5	&5	&107	&173&	70\\\hline
-4	&125&	13	&93	&197&	246\\\hline
-3	&225&	35	&122	&161&	239\\\hline
-2	&342&	39	&156	&202	&195\\\hline
-1	&214&	16	&111	&104	&175\\\hline
0	&10	&18&	93&	101&	102\\\hline

\end{tabular}
 \caption{Quarter 1 performance with respect to Class Lag} 
    \label{class_lag}
\end{table}

Since it will not be a desirable idea to grade the students with varied class lags on the same basis, we establish a score measure which gives a unique score for improvement for students with varied class lag. Also to be noted is that the level of difficulty of the students and assessments may vary 
from quarter to quarter. Therefore the scores for the improvement also reflect that by considering the overall performance.

\subsubsection*{Scores for improvement measurement}

For each quarter, depending on the overall performance of all the students a score may be used for measurement of improvement, given the existing class lag of the student.\\ 
Steps for calculating the score: 
\begin{enumerate}  
  \item 	Compute a cross tabulation of class tab and  number of subjects showing improvement, like Table 1. 
  \item 	Calculate the row sum, which is the number of students with that particular class lag.
  \item 	Divide the row by the row sum in order to get the proportion of students showing improvement in number of subjects for each class lag. 
  \item 	Compute the cumulative proportion and assign 0 to the column improvement in 0 subject in order to obtain the desired score.
\end{enumerate}

As an illustration for Quarter 1 performance (Table \ref{class_lag}), the score table will be as follows: 
\begin{table}[h]
    \centering
     \rowcolors{2}{blue!12}{white}
\begin{tabular}{|c|c|c|c|c|c|}\hline
\backslashbox{Class lag}{Improvement}
&\makebox[1.5em]{0}&\makebox[1.5em]{1}&\makebox[1.5em]{2}
&\makebox[1.5em]{3}&\makebox[1.5em]{4}\\\hline\hline
-7&	0.00&	0.06&	0.39	&0.95&	1.00\\\hline
-6	&0.00	&0.03&	0.27&	0.88&	1.00\\\hline
-5	&0.00	&0.03&	0.33&	0.81&	1.00\\\hline
-4	&0.00	&0.20&	0.34&	0.64&	1.00\\\hline
-3	&0.00	&0.33&	0.49&	0.69&	1.00\\\hline
-2	&0.00	&0.41&	0.57&	0.79&	1.00\\\hline
-1	&0.00	&0.37&	0.55&	0.72&	1.00\\\hline
0	&0.00	&0.09&	0.37&	0.69&	1.00\\\hline
\end{tabular}
 \caption{Quarter 1 performance with respect to Class Lag} 
    \label{class_lag_scores}
\end{table}

\subsubsection*{Interpretation} 

The value of the score lies between 0 and 1.  Naturally the score rises as the number of subjects with improvement increases. The score 1 indicates improvement in all 4 subjects while score 0 indicates improvement in none of the subjects. 
As can be observed from Table 1.4; in quarter 1, a student with class lag of 7 and showing improvement in exactly 2 subjects obtains a score 0.39 while if he improves in 3 subjects the score rises to 0.95.  Similarly for a student with no class lag and showing improvement in 3 subjects the score is 0.69. 

\subsection{Analysis of Progression}

This section corresponds to measuring progression in terms of improvement performance of the students over the 3 quarters. For this propose, a progression score measure has been used considering the improvement level in both the quarters.

\subsubsection*{Calculation of Progression Score}

A progression rate matrix is calculated for inter-quarter improvement comparison as follows:
\begin{enumerate}
    \item Compute a cross tabulation of improvement level (0,1,2,3,4) with rows for first quarter and columns for second quarter. 
\item Calculate the row sum, which is the number of students with that improvement level in quarter 1. 
\item Divide the row by the row sum in order to get the proportion of students showing the increase/decrease in improvement level.
\end{enumerate}
Define  $ S= \sum_{i=0}^{4}\sum_{i=0}^{4}p_{ij} (j-i) $, where $p_{ij}$  are the entries of the Progression-rate matrix
which, after simplification becomes: $S= \sum_{j=0}^{4}jp_{.j}-10$
Next scaling S so as to define the \textbf{score value $S^*$}, which gives the progression score lying between 0 to 1. 
A negative score indicates lack of improvement over the quarter while a positive score indicates degree of improvement over the quarters. 

\subsubsection*{Calculation of Progression Score from Quarter 1 to Quarter 2}

\textbf{Step 1: } Cross Tabulation 
\begin{table}[h]
    \centering
     \rowcolors{2}{blue!12}{white}
\begin{tabular}{|c|c|c|c|c|}\hline
\backslashbox{Quarter 1}{Quarter 2}
&\makebox[1.5em]{1}&\makebox[1.5em]{2}
&\makebox[1.5em]{3}&\makebox[1.5em]{4}\\\hline\hline
0	&48&	83	&47&	754\\ \hline
1	&21	&24	&19&	67\\ \hline
2	&25	&173&	161&	399\\ \hline
3	&24	&208&283&597\\ \hline
4	&22	&178&255&	612\\ \hline

\end{tabular}
\end{table}

\textbf{Step 2: } Calculating Row Totals
\begin{table}[h]
    \centering
     \rowcolors{2}{blue!12}{white}
\begin{tabular}{|c|c|c|c|c|c|}\hline
\backslashbox{Quarter 1}{Quarter 2}
&\makebox[1.4em]{1}&\makebox[1.4em]{2}
&\makebox[1.4em]{3}&\makebox[1.4em]{4}&\makebox[3.3em]{Row Sum}\\\hline\hline
0	&48&	83	&47&	754&932\\ \hline
1	&21	&24	&19&	67&131\\ \hline
2	&25	&173&	161&	399&758\\ \hline
3	&24	&208&283&597&1112\\ \hline
4	&22	&178&255&	612&1067\\ \hline
\end{tabular}
\end{table}

\textbf{Step 3: }Progression Rate Matrix
\begin{table}[h]
    \centering
     \rowcolors{2}{blue!12}{white}
\begin{tabular}{|c|c|c|c|c|c|}\hline
\backslashbox{Quarter 1}{Quarter 2}
&\makebox[1.4em]{1}&\makebox[1.4em]{2}
&\makebox[1.4em]{3}&\makebox[1.4em]{4}\\\hline\hline
0	&5.15\%	&8.91\%	&5.04\%&	80.90\%\\\hline
1	&16.03\%&18.32\%&14.50\%&51.15\%\\\hline
2	&3.30\%	&22.82\%&	21.24\%&52.64\%\\\hline
3	&2.16\%	&18.71\%&	25.45\%	&53.69\%\\\hline
4	&2.06\%	&16.68\%&	23.90\%	&57.36\%\\\hline

\rowcolor{gray!50}
Column sum&	28.70\%&	85.44\%&	90.14\%&	295.73\%
\\\hline

\end{tabular}
\end{table}
\FloatBarrier
Now, $S= \sum_{j=1}^{4}jp_{.j}-10 = 6.52$, and $S^* = \frac{S}{30}=0.217$.
The Progression Score is 0.217 which indicates an overall progression from Quarter 1 to Quarter 2.

\subsubsection{Analysis over the 3 Quarters}

Comparing the performance in terms of the number of subjects in which a student shows improvement over the 3 Quarters, the following grading system has been used corresponding to the improvement level. 
\begin{figure}[h!]
    \centering
    \includegraphics[width=1\textwidth]{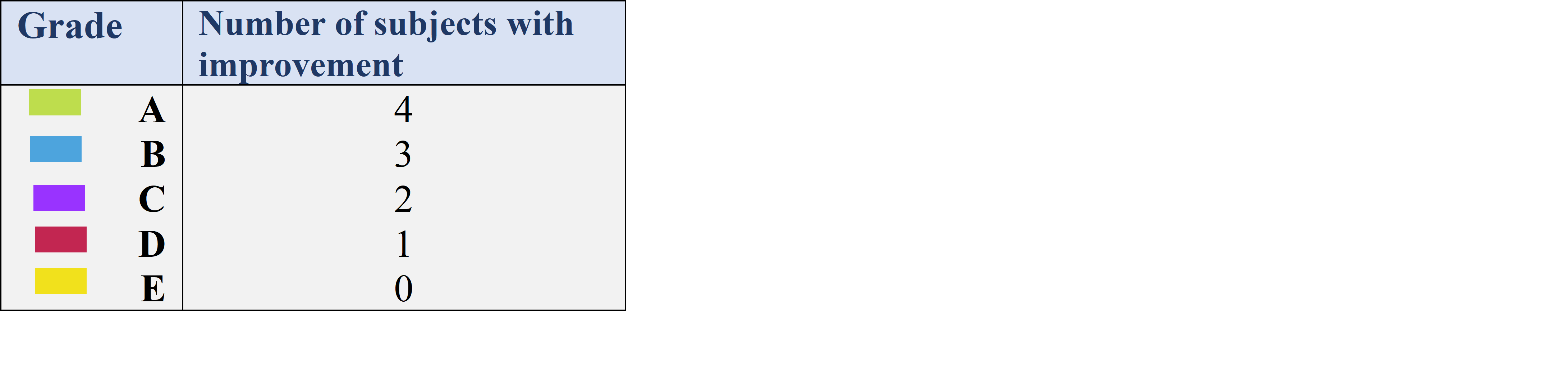}
\end{figure}
\FloatBarrier
\begin{figure}[h!]
    \centering
    \includegraphics[width=1\textwidth]{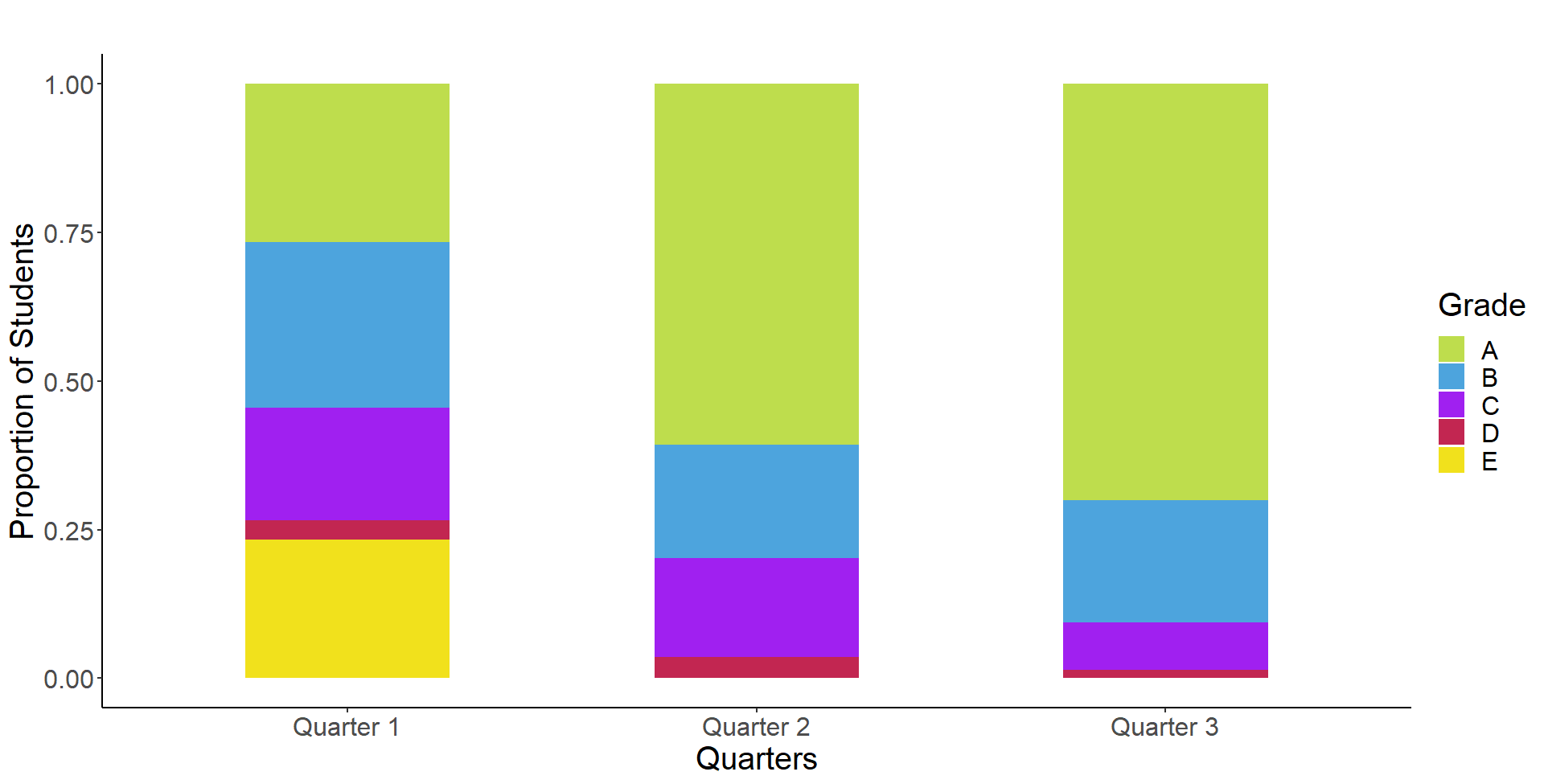}
    \caption{Figure showing performance over the 3 Quarters}
    \label{3qrt_overall}
\end{figure}
It can be observed that proportion of students securing Grade A i.e., showing improvement in all 4 subjects has increased over the 3 semesters (from 26.67\% in Quarter 1 to 60.725\% in Quarter 2 to 70.1\% in Quarter 3) . None of the student secured Grade E (for showing no improvement) in Quarter 2 and Quarter 3. \\
Using similar method as before, the progression score for the quarters 2 and 3 has been calculated (Refer to \hyperref[Appendix-progressionmatrix]{Appendix A}).  \\
The progression score from Quarter 2 to Quarter 3 is 0.122. We observe that overall progression has been positive over the 3 quarters. However the progression from quarter 1 to quarter 2 is higher compared that from quarter 2 to quarter 3. The level of difficulty of assessments may have increased over the time period leading to decrease in progression. 

\subsubsection{Comparing Male-Female Progression }

Of the 4000 students of which 2119 (53\%) are females and 1881(47\%) are males. 
\begin{figure}[h]
    \centering
    \begin{subfigure}[b]{0.6\textwidth}
         \centering
         \includegraphics[width=0.7\textwidth]{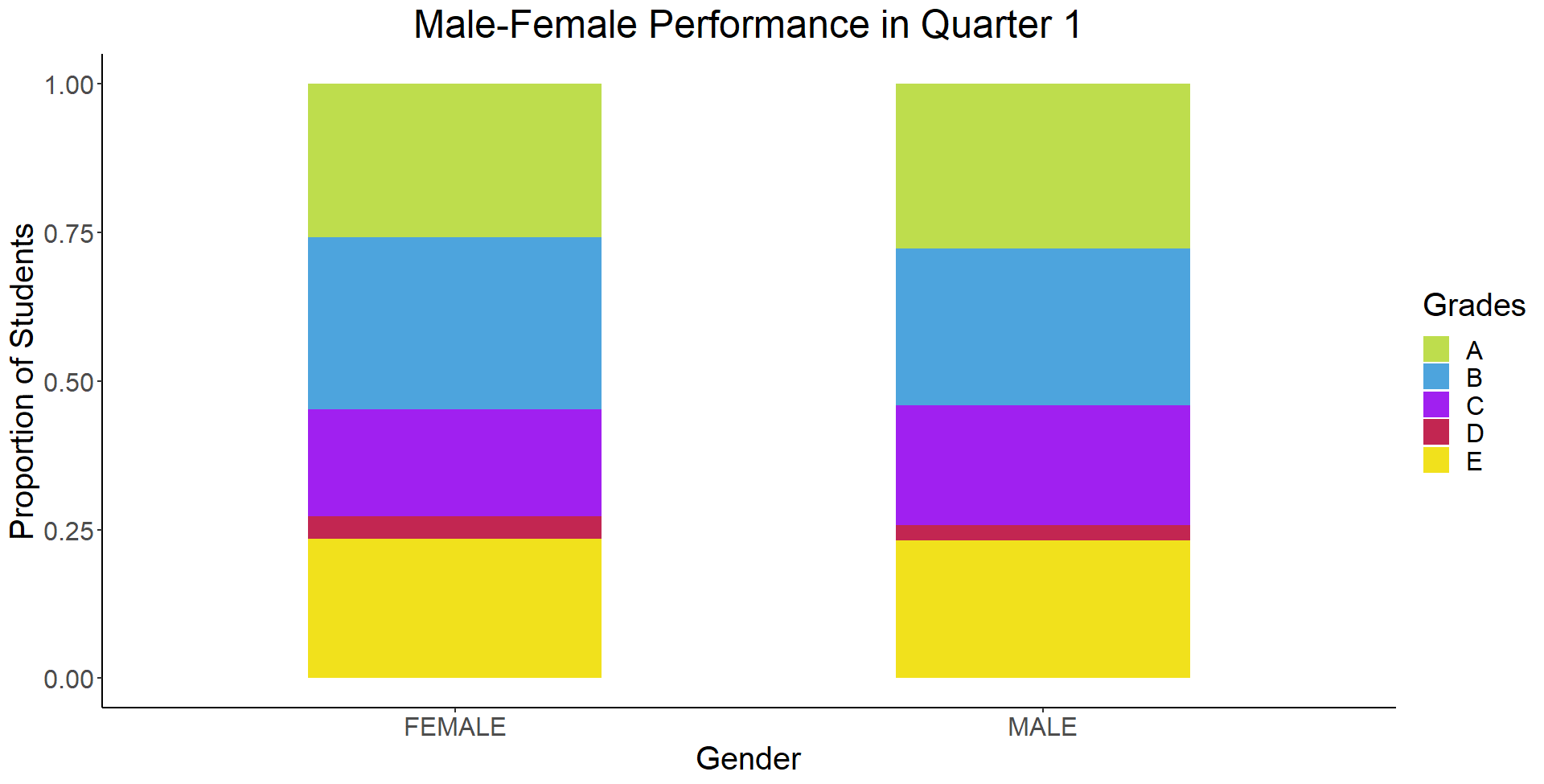} 
     \end{subfigure}
     \vfill
     \begin{subfigure}[b]{0.6\textwidth}
         \centering
         \includegraphics[width=0.7\textwidth]{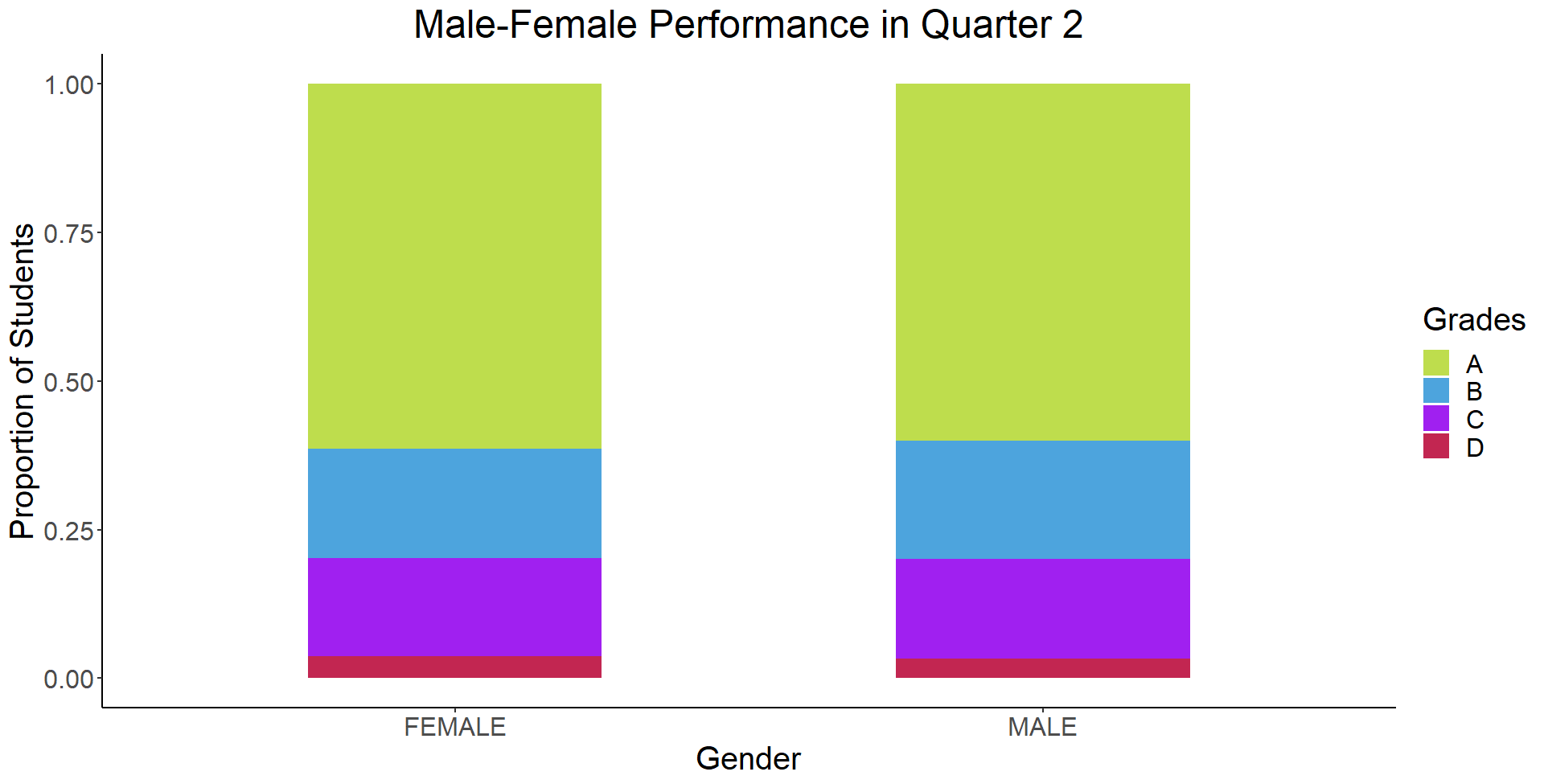}   
     \end{subfigure}
     \vfill
     \begin{subfigure}[b]{0.6\textwidth}
         \centering
         \includegraphics[width=0.7\textwidth]{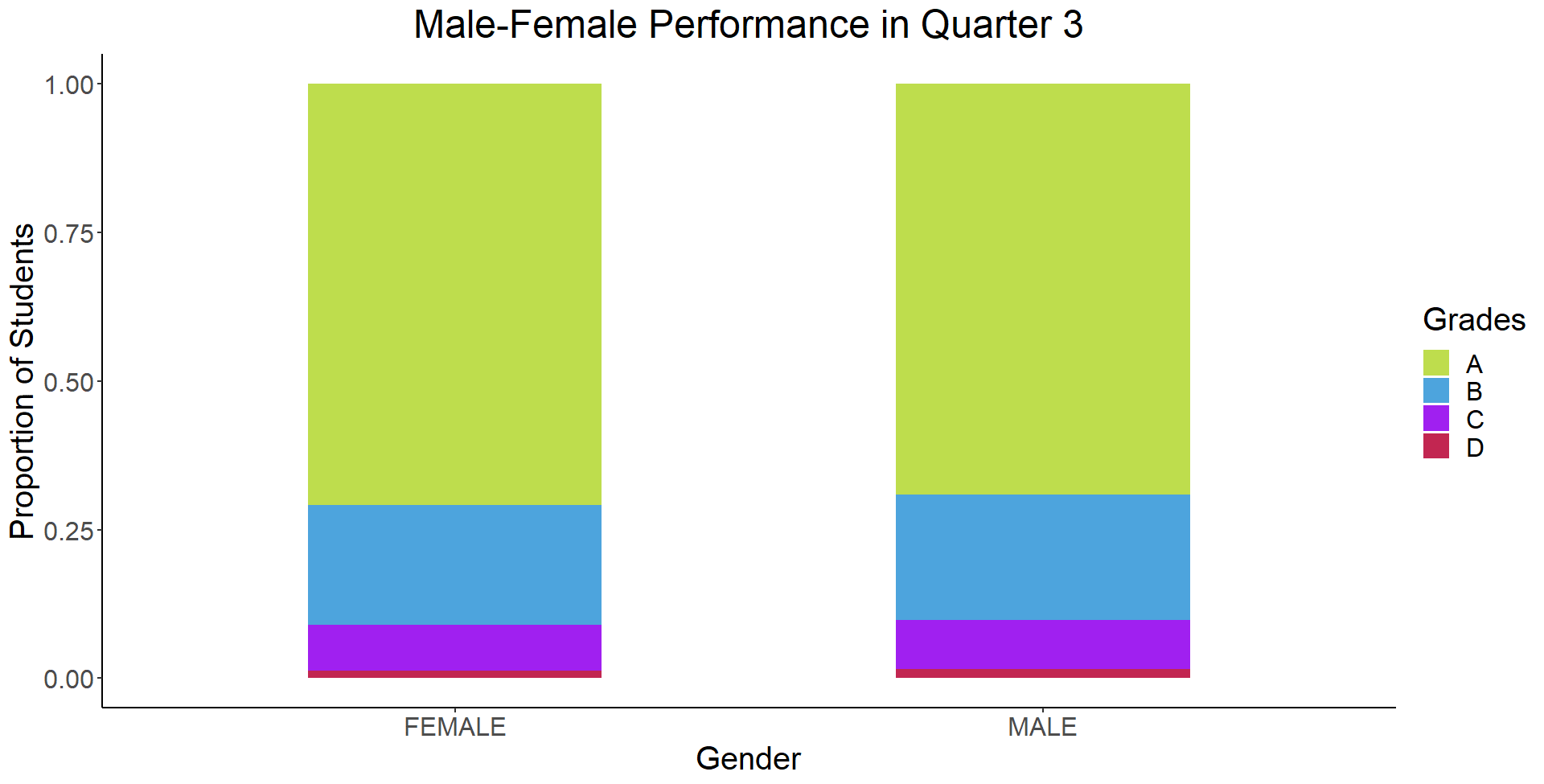}   
     \end{subfigure}
    \caption{Figures showing representation of Male-Female performance over the 3 Quarters}
    
\end{figure}
Plotting male-female performance over the 3 Quarters we observe almost similar proportion as per gender of the students among the various Grades. \\
Calculating Progression scores as before, we find that: 
\begin{center}
\begin{tabular}{ | m{3em} | m{5cm}| m{5cm} | } 
  \hline
   \rowcolor{blue!23}
  Gender&	Quarter1 to Quarter2&	Quarter2 to Quarter3 \\ 
  \hline
 Female	&0.217&	0.125 \\ 
  \hline
 Male&	0.218	&0.119 \\ 
  \hline
\end{tabular}
\end{center}
It was observed that from Quarter 2 to Quarter 1 the progression score of females and males have been same approximately. However from Quarter 2 to Quarter 3 females show higher progression compared to males. The progression rate in Quarter 2 to Quarter 3 is less than that in Quarter 1 to Quarter 2 for both males and females. 

\subsubsection{Comparing State-wise Progression }

\begin{figure}[h]
    \centering
    \begin{subfigure}[b]{0.6\textwidth}
         \centering
         \includegraphics[width=0.95\textwidth]{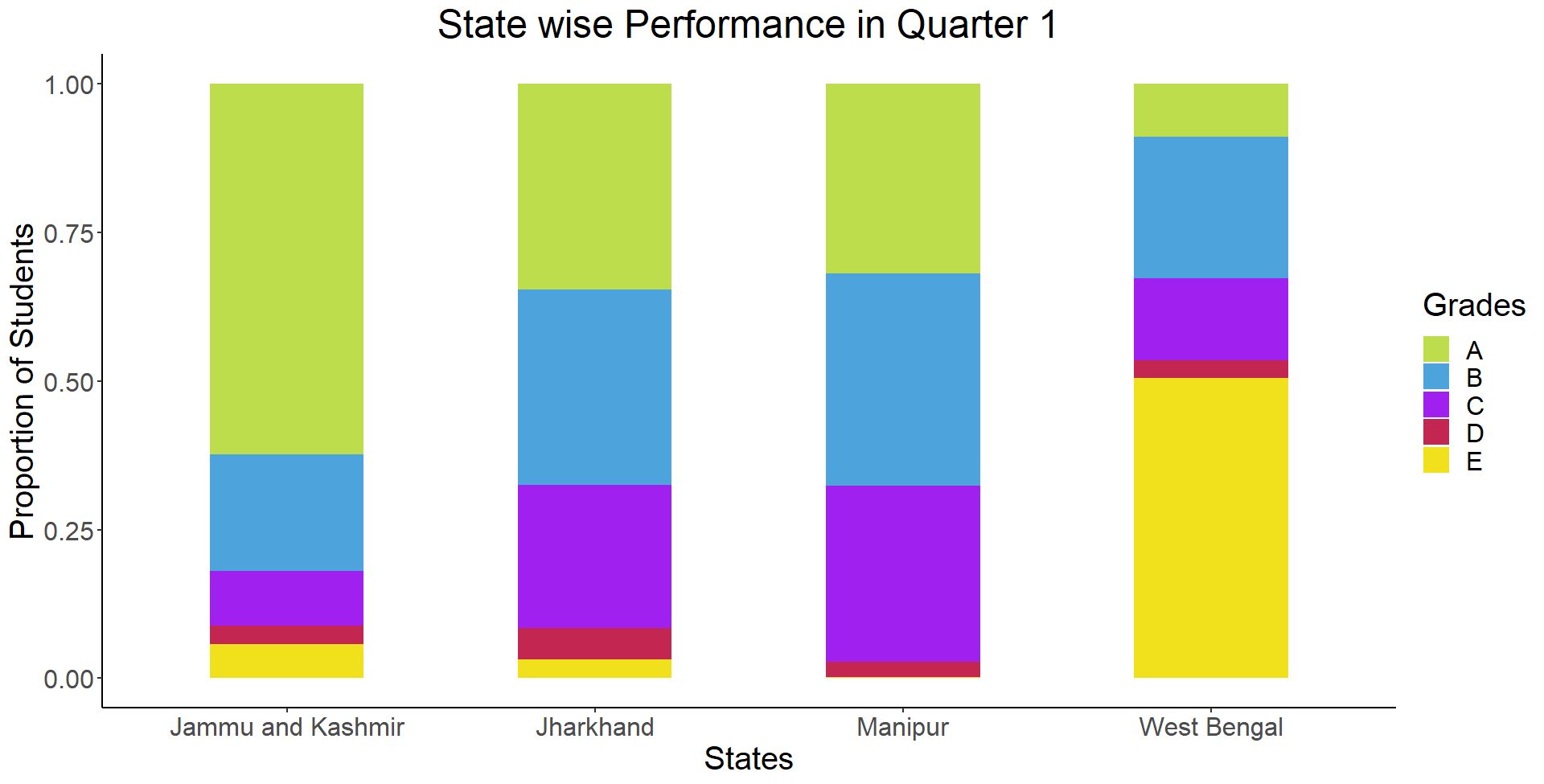} 
     \end{subfigure}
     \vfill
     \begin{subfigure}[b]{0.6\textwidth}
         \centering
         \includegraphics[width=0.95\textwidth]{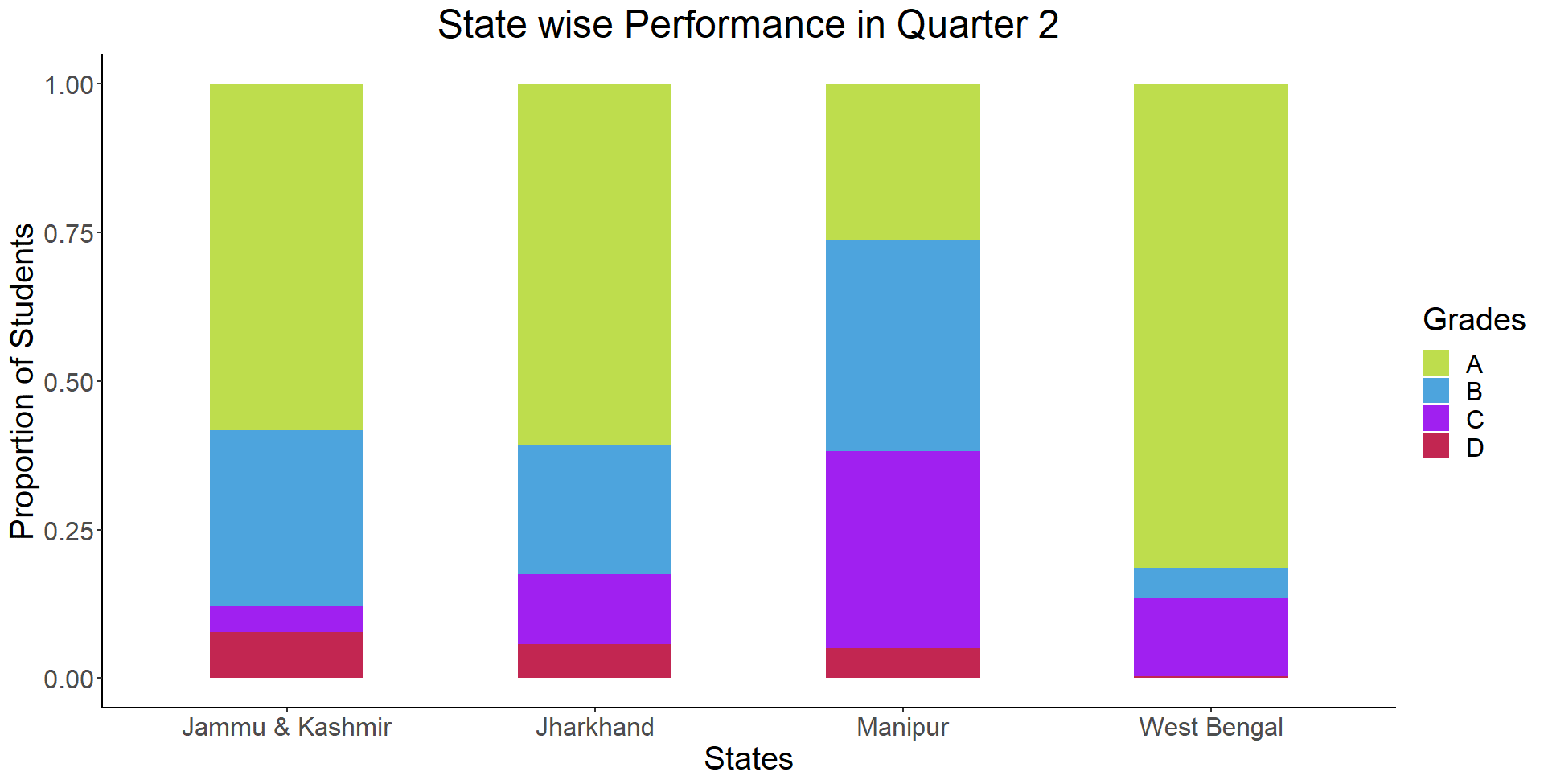}   
     \end{subfigure}
     \vfill
     \begin{subfigure}[b]{0.6\textwidth}
         \centering
         \includegraphics[width=0.95\textwidth]{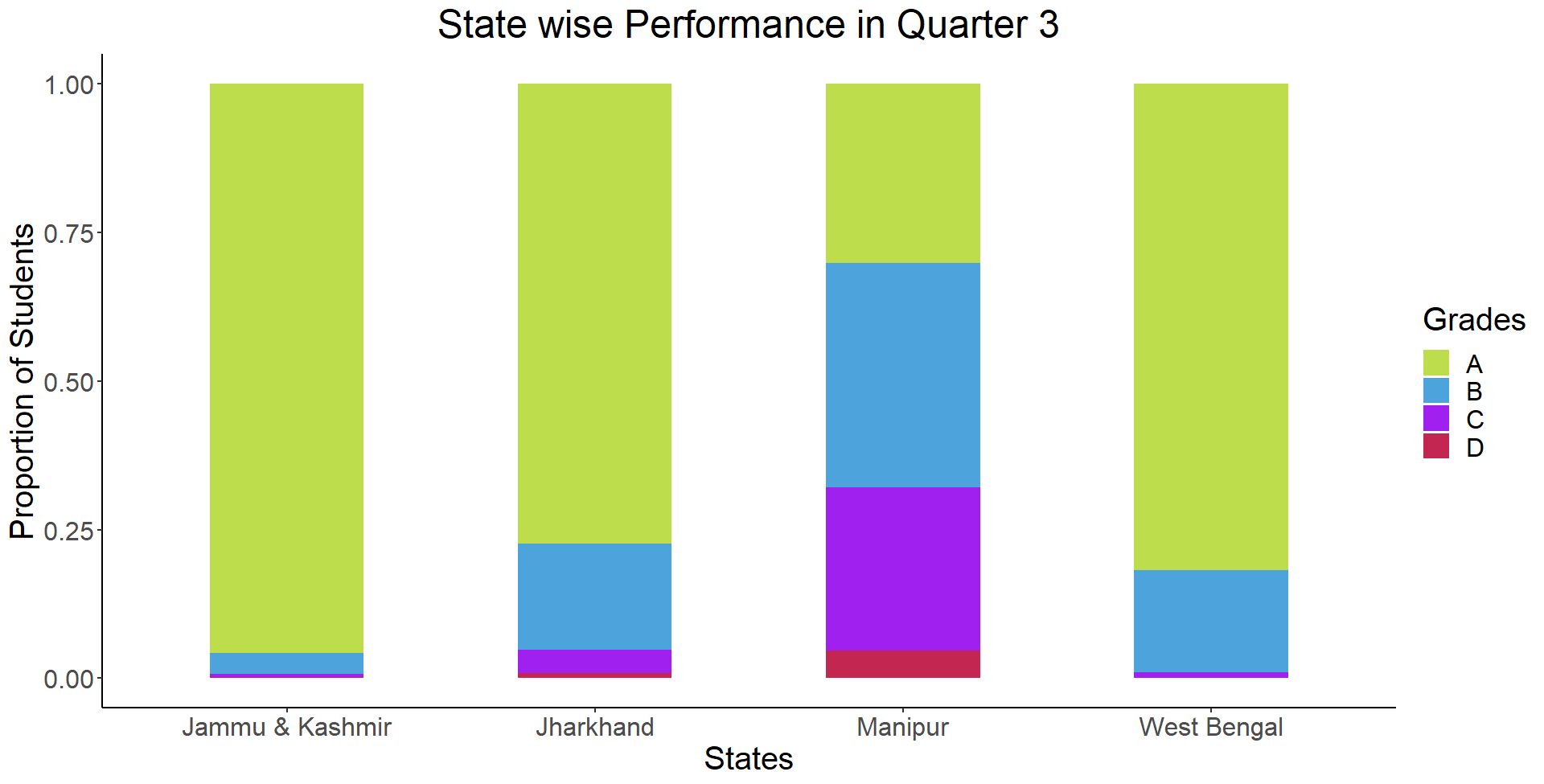}   
     \end{subfigure}
    \caption{Figures showing representation of Inter-state performance over the 3 Quarters}
    \label{state-wise}
\end{figure}
Plotting the state wise performance over the 3 quarters, it was observed that the proportion of students securing Grade A has increased from Quarter 1 to Quarter 2 in Jammu and Kashmir, Jharkhand and West Bengal. The performance of Manipur has been similar over the 3 Quarters. A clear progression can be seen in West Bengal centres where in Quarter 1 only 8.89\% students got Grade A while in Quarter 2,  81.37\% got Grade A. From Quarter 2 to Quarter 3, the percentage of students securing Grade A increased from 58.2\% to 95.7\% in Jammu and Kashmir. In Quarter 2 and Quarter 3 none of the student got Grade E.\\
Calculating Progression scores as before, we find that: 
\begin{center}
\begin{tabular}{ | m{9em} | m{4cm}| m{4cm} | } 
  \hline
   \rowcolor{blue!23}
  States&	Quarter1 to Quarter2&	Quarter2 to Quarter3 \\ 
  \hline
Jammu \& Kashmir	&0.096&	0.187\\ \hline
Jharkhand	&0.171&	0.163\\ \hline
Manipur	&0.105	&0.042\\ \hline
West Bengal	&0.260&	0.155\\ \hline

\end{tabular}
\end{center}
It can be observed from Quarter 1 to Quarter 2, West Bengal has shown the maximum progression compared to the other states, followed by Jharkhand and from Quarter 2 to Quarter 3, Jammu \& Kashmir has shown the maximum progression compared to the other states. The progression scores have increased in Quarter 2 to Quarter 3 performance for Jammu \& Kashmir. Manipur observes the lowest progression, which also follows from Figure \ref{state-wise}. 

\section{Conclusion}
%A conclusion section is not required. Although a conclusion may review the  main points of the paper, do not replicate the abstract as the conclusion. A  conclusion might elaborate on the importance of the work or suggest  applications and extensions.
Globally, the COVID-19 pandemic had affected the educational system. To maintain social distance, the
pandemic prompted an immediate and total lockdown of all educational institutions. The lockdown seems to have generally caused a severe impact on the learning of students
and a shift in their learning methods. These students were not able to learn on a one-on-one basis with their educators, as the pandemic initiated an immediate and complete close-down of all educational institutions, the shift in learning from traditional classroom learning to computer-based learning became one of the greatest academic changes that the students needed to cope with. The findings
of this assessment help in developing a better understanding of required educational reforms in the pandemic
and post-pandemic times, as the education system needs to be transformed significantly instead of waiting for normalcy.

CRY and RILM initiated a project to assess the learning acievement of 4000 children across four states: Jammu \& Kashmir, Jharkhand, Manipur and West Bengal. Every child was provided with the books of appropriate class according to their age in order to test their competency in reading and basic calculation
as well and thereafter the compatible class was determined. The assessments were carried  over 3 quarters in 4 subjects: oral assessments in $1^{st}$ Language, $2^{nd}$, 
Language and  Mathematics and a writing assessment and a binary variable for improvement/no improvement (1/0) was provided. 

For this analysis, we proposed a measure which gives a unique score for improvement level of students with varied class lag. Based on data from the four states, the learning progression of all 4000 students over the 3 quarters of an academic year have been studied. It was observed that at all levels progression score is positive, showing improvement from quarter to quarter performance in terms of making up the academic lag.
The analysis of progression has been carried out by gender and state level for male-female and inter state comparison respectively.

%\section*{Acknowledgment}
%We acknowledge CRY for providing us with the data sets containing details of the child ID, their age appropriate class, comparison class and assessment scores for the 3 quarters of the students.
\newpage
\section*{References}
\begingroup
\renewcommand{\section}[2]{}%

\endgroup

%\newpage

%\newpage
\section*{Appendix A}
\label{Appendix-progressionmatrix}
\subsection{Calculation of Progression Score from Quarter 2 to 3}
\textbf{Step 1: } Cross Tabulation 
\begin{table}[h]
    \centering
     \rowcolors{2}{blue!12}{white}
\begin{tabular}{|c|c|c|c|c|}\hline
\backslashbox{Quarter 2}{Quarter 3}
&\makebox[1.5em]{1}&\makebox[1.5em]{2}
&\makebox[1.5em]{3}&\makebox[1.5em]{4}\\\hline\hline
1	&13&	18&	16&	93\\ \hline
2	&13&	150	&239&	264\\ \hline
3	&9	&77&	267&	412\\ \hline
4	&19&	76&	299&	2035\\ \hline
\end{tabular}
\end{table}
\\

\textbf{Step 2: }Progression Rate Matrix
\begin{table}[h]
    \centering
     \rowcolors{2}{blue!12}{white}
\begin{tabular}{|c|c|c|c|c|c|}\hline
\backslashbox{Quarter 1}{Quarter 2}
&\makebox[1.4em]{1}&\makebox[1.4em]{2}
&\makebox[1.4em]{3}&\makebox[1.4em]{4}\\\hline\hline
1	&9.29\%	&12.86\%&	11.43\%&	66.43\%\\\hline
2	&1.95\%	&22.52\%&	35.89\%&	39.64\%\\\hline
3	&1.18\%	&10.07\%&	34.90\%&	53.86\%\\\hline
4	&0.78\%	&3.13\%	&12.31\%&	83.78\%\\\hline

\rowcolor{gray!50}
Column sum&	13.19\%&	48.57\%&	94.52\%&	243.708\%
\\\hline

\end{tabular}
\end{table}

Now, $S= \sum_{j=1}^{4}jp_{.j}-10 = 3.68$, and $S^* = \frac{S}{30}=0.122$.
The Progression Score is 0.122 which indicates an overall progression from Quarter 2 to Quarter 3.

\end{document}